# Injury Severity Analysis of Truck-Involved Crashes under Different Weather Conditions

Majbah Uddin[a] and Nathan Huynh[b]*

[a]Oak Ridge National Laboratory
National Transportation Research Center
2360 Cherahala Blvd, Knoxville, TN 37932, USA

[b]University of South Carolina
Department of Civil and Environmental Engineering
300 Main St, Columbia, SC 29208, USA

*E-mail addresses*: uddinm@ornl.gov (M. Uddin), nathan.huynh@sc.edu (N. Huynh)

*Corresponding Author Contact Information
Nathan Huynh
University of South Carolina
Department of Civil and Environmental Engineering
300 Main St, Columbia, SC 29208, USA
Telephone: (803) 777-8947
Fax: (803) 777-0670
Email: nathan.huynh@sc.edu

**Abstract**

This paper investigates truck-involved crashes to determine the statistically significant factors that contribute to injury severity under different weather conditions. The analysis uses crash data from the state of Ohio between 2011 and 2015 available from the Highway Safety Information System. To determine if weather conditions should be considered separately for truck safety analyses, parameter transferability tests are conducted; the results suggest that weather conditions should be modeled separately with a high level of statistical confidence. To this end, three separate mixed logit models are estimated for three different weather conditions: normal, rain and snow. The estimated models identify a variety of statistically significant factors influencing the injury severity. Different weather conditions are found to have different contributing effects on injury severity in truck-involved crashes. Rural, rear-end and sideswipe crash parameters were found to have significantly different levels of impact on injury severity. Based on the findings of this study, several countermeasures are suggested: 1) safety and enforcement programs should focus on female truck drivers, 2) a variable speed limit sign should be used to lower speeds of trucks during rainy condition, and 3) trucks should be restricted or prohibited on non-interstates during rainy and snowy conditions. These countermeasures could reduce the number and severity of truck-involved crashes under different weather conditions.

## 1. Introduction

Interest in identifying factors that affect truck transportation safety in the U.S. has increased in recent years due to the higher number of fatalities from truck-involved crashes, a byproduct of the growing domestic e-commerce and international trade (Ahmed et al., 2018; Al-Bdairi and Hernandez, 2017; Cerwick et al., 2014; Chang and Chien, 2013; Chen and Chen, 2011; Islam et al., 2014; Islam and Hernandez, 2013a,b; Islam and Ozkul, 2019; Lyman and Braver, 2003; Uddin and Huynh, 2017, 2018; Zaloshnja and Miller, 2004). In 2015, there were 32,166 fatal crashes on U.S. roadways, of which, 3,598 (11.2%) involved at least one truck. The number of fatalities in the U.S. when a truck is involved in a crash in 2015 during inclement weather, such as rain, snow, sleet, hail, fog, and severe crosswinds was 458 (Federal Motor Carrier Safety Administration, 2017). Compared to passenger vehicles, trucks are more vulnerable to crashes in inclement weather due to their larger size and higher center of gravity. At the state level, Ohio had a very high number of fatal truck-involved crashes (156) in 2015 (Federal Motor Carrier Safety Administration, 2017).

This study is focused on investigating the relationship between crash factors and crash injury severity, based on different weather conditions which have not been studied previously. Past studies have indicated that roadway weather conditions play a significant role in injury severity from truck-involved crashes (e.g., Anderson and Hernandez, 2017; Cerwick et al., 2014; Chen and Chen, 2011; Dong et al., 2015; Islam et al., 2014; Islam and Hernandez, 2013b; Khorashadi et al., 2003; Lemp et al., 2011; Li et al., 2017; Naik et al., 2016; Osman et al., 2016; Pahukula et al., 2015; Uddin and Huynh, 2017, 2018). However, these studies have not examined the impact of weather conditions via separate models for different weather conditions. The interaction between variables is complex, which can vary significantly across different weather conditions. For instance, while the aggregate model may indicate that daylight decreases injury severity of truck drivers, its effect may vary under different weather conditions. That is, the injury severity of drivers may be less severe under daylight and rainy conditions (Pahukula et al., 2015), but more severe under daylight and snowy conditions (Forkenbrock and Hanley, 2003). As such, disaggregating truck-involved crashes by weather conditions can provide additional insights to traffic safety engineers and transportation planners about the effect of weather conditions on truck-involved crashes, and thereby, enabling them to implement appropriate countermeasures. Furthermore, in recent years more and more researchers have adopted the use of separate models in analyzing truck-involved crashes: rural and urban (Chen and Chen, 2011; Islam et al., 2014), time of day (Behnood and Mannering, 2019; Pahukula et al., 2015), roadway classification (Anderson and Hernandez, 2017), and lighting condition (Uddin and Huynh, 2017).

As for methodology, most of the previous studies that examined truck-involved crashes modeled injury severity using logit or probit models (e.g., Al-Bdairi et al., 2017; Behnood and Mannering, 2019; Cerwick et al., 2014; Chen and Chen, 2011; Duncan et al., 1998; Islam and Hernandez, 2013a,b; Islam et al., 2014; Islam 2015; Khorashadi et al., 2005; Lemp et al., 2011; Naik et al., 2016; Pahukula et al., 2015; Taylor et al., 2017; Uddin and Huynh, 2017, 2018; Wei et al., 2017). Some of these studies considered the injury severity of the driver as the dependent variable while others considered the injury severity of the most severely injured occupant. In this study, the injury severity of the truck driver is chosen to be the dependent variable. Furthermore, mixed logit (random parameters logit) modeling is used to determine the contributing factors and to account for the unobserved heterogeneity. Mixed logit models are statistically superior to traditional fixed parameters logit models and they require less detailed crash-specific data than that of fixed parameters models (Anastasopoulos and Mannering, 2011).

The objective of this study is to investigate the factors that influence injury severity of drivers from truck-involved crashes under three different weather conditions (at the time of the crash): normal, rain and snow. The analysis uses crash data from the state of Ohio between 2011 and 2015 available from the Highway Safety Information System (HSIS). To the best of the authors' knowledge, this study is the first to analyze driver injury severity in truck-involved crashes under different weather conditions.

## 2. Previous research

A number of studies have explored injury severity of truck-involved crashes. The research topics include determining contributing crash factors, interactions between the factors, and comparison of methodologies. Readers are referred to the review paper by Savolainen et al. (2011) for more information about these research topics. Research on the effect of weather conditions on driver injury severity in truck-involved crashes is limited. Young and Liesman (2007) used 1994 to 2003 Wyoming truck crash data to examine the relationship between wind speed and truck overturning via a binary logit model. Their modeling results indicated that wind speed could be used as a predictor of truck overturning in a crash. However, their study did not explore the effect of wind speed on injury severity. Kecojevic and Radomsky (2004) used 1995 to 2002 fatal crash data from the Mine Safety and Health Administration and found that inclement weather conditions and truck-involved crashes are related. The authors performed percentage analysis to determine the impact of different crash types and crash reasons. Naik et al. (2016) investigated truck crash injury severity in Nebraska using an aggregated data set (15-minute weather station data combined with crash and roadway data) from 2009 to 2011. The authors used both ordered and multinomial logit models. They found that inclement weather conditions had an effect on truck-involved crash injury severity. Specifically, the greater the recorded wind speed and rain, the more severe the injury in crashes.

The aforementioned studies indicated that weather conditions have a significant impact on truck-involved crash injury severity; however, they have not examined how the factors contribute to the injury severity under different weather conditions. This study aims to fill this gap in the literature by developing a mixed logit model for each type of weather condition.

## 3. Data description

The data used in this study are highway patrol reported crashes from the state of Ohio between 2011 and 2015, available from the Highway Safety Information System (HSIS) database. Using the vehicle type attribute, crash data were filtered to include only crashes involving trucks. Specifically, only crashes involving single-unit trucks, truck trailers, tractor semi-trailers and tractor doubles were considered. Note that both at-fault and no-fault (i.e., non-contributing) truck-involved crashes are included in the dataset. Also, only crashes which occurred along roadway segments were considered. That is, intersection crashes were excluded. The reason is because factors that affect crashes along segments and crashes at intersections are significantly different according to Vogt and Bared (1998). Therefore, to properly capture the impact of location type, segment and intersection crashes need to be modeled separately. Furthermore, in the U.S., there were a larger number of fatal (2,649) and injury (50,000) truck-involved crashes that occurred along roadway segments in 2015 than at intersections (Federal Motor Carrier Safety Administration, 2017).

The resulting dataset has three weather conditions: normal, rain and snow. These three weather conditions were considered due to their sample shares. Other conditions such as fog and heavy wind had very low sample shares, and thus, not sufficient for model development. Each observation in the dataset includes the injury severity of the driver of the truck along with driver, crash, vehicle, roadway and temporal characteristics.

The final dataset consists of 49,248 truck-involved crashes. Of this total, 40,459 occurred during normal condition, 4,866 occurred during rainy condition and 3,923 occurred during snowy condition. The injury severity of the crash data in the HSIS database is categorized into five distinct levels: fatal (105 or 0.2%), disabling injury (424 or 0.9%), evident injury (3,328 or 6.8%), possible injury (1,665 or 3.4%) and no injury (43,726 or 88.7%). As done in other studies (Chen and Chen, 2011; Islam et al., 2014; Uddin and Ahmed, 2018; Uddin and Huynh, 2017, 2018), to ensure sufficient number of observations for each injury severity level, the above five injury severity levels were consolidated into three levels: major injury (fatality and disabling injury), minor injury (evident injury and possible injury) and no injury. Table 1 presents the injury severity level frequency and percentage distribution by weather conditions.

**Table 1**
Injury severity level frequency and percentage distribution by weather conditions.

| Weather condition | Total observation | Major injury (%) | Minor injury (%) | No injury (%) |
| --- | --- | --- | --- | --- |
| Normal | 40,459 | 443 (1.1) | 4,023 (9.9) | 35,993 (89.0) |
| Rain | 4,866 | 47 (1.0) | 511 (10.5) | 4,308 (88.5) |
| Snow | 3,923 | 39 (1.0) | 459 (11.7) | 3,425 (87.3) |

**Table 2**
Descriptive statistics of variables by weather conditions.

| Meaning of variable | Normal | | Rain | | Snow | |
| --- | --- | --- | --- | --- | --- | --- |
| | Mean | SD† | Mean | SD† | Mean | SD† |
| *Driver characteristics* | | | | | | |
| Male (1 if male driver, 0 otherwise) | 0.96 | 0.20 | 0.96 | 0.20 | 0.96 | 0.20 |
| Restraint (1 if used lap and/or shoulder belt, 0 otherwise) | 0.94 | 0.23 | 0.94 | 0.23 | 0.95 | 0.21 |
| | | | | | | |
| *Crash characteristics* | | | | | | |
| Rural (1 if rural location, 0 otherwise) | 0.38 | 0.48 | 0.34 | 0.47 | 0.48 | 0.50 |
| Urban (1 if urban location, 0 otherwise) | 0.62 | 0.48 | 0.66 | 0.47 | 0.52 | 0.50 |
| Curve (1 if curved highway, 0 otherwise) | 0.10 | 0.30 | 0.15 | 0.36 | 0.11 | 0.32 |
| Rear-end (1 if rear-end collision, 0 otherwise) | 0.19 | 0.39 | 0.19 | 0.39 | 0.21 | 0.41 |
| Sideswipe (1 if sideswipe collision, 0 otherwise) | 0.32 | 0.47 | 0.32 | 0.47 | 0.34 | 0.47 |
| Object (1 if collision with an object, 0 otherwise) | 0.14 | 0.35 | 0.20 | 0.40 | 0.19 | 0.39 |
| MVIT (1 if collision with a motor vehicle in transport, 0 otherwise) | 0.63 | 0.48 | 0.64 | 0.48 | 0.66 | 0.48 |
| Ran off (1 if ran off road to the right or left, 0 otherwise) | 0.10 | 0.30 | 0.16 | 0.37 | 0.18 | 0.38 |
| Daylight (1 if daylight, 0 otherwise) | 0.77 | 0.42 | 0.65 | 0.48 | 0.61 | 0.49 |
| Dark-lighted (1 if dark with streetlights, 0 otherwise) | 0.08 | 0.27 | 0.15 | 0.36 | 0.12 | 0.33 |
| Dark-unlighted (1 if dark without streetlights, 0 otherwise) | 0.13 | 0.34 | 0.18 | 0.38 | 0.25 | 0.43 |
| | | | | | | |
| *Vehicle characteristics* | | | | | | |
| Single-unit truck (1 if single-unit truck, 0 otherwise) | 0.28 | 0.45 | 0.26 | 0.44 | 0.25 | 0.43 |
| Truck trailer (1 if truck trailer, 0 otherwise) | 0.11 | 0.31 | 0.11 | 0.31 | 0.08 | 0.26 |
| Truck semi-trailer (1 if truck semi-trailer, 0 otherwise) | 0.59 | 0.49 | 0.61 | 0.49 | 0.64 | 0.48 |
| | | | | | | |
| *Roadway characteristics* | | | | | | |
| Speed1 (1 if speed limit ≤ 40 mph, 0 otherwise) | 0.21 | 0.41 | 0.20 | 0.40 | 0.13 | 0.34 |
| Speed2 (1 if speed limit 45 mph–60 mph, 0 otherwise) | 0.38 | 0.49 | 0.37 | 0.48 | 0.31 | 0.46 |
| Speed3 (1 if speed limit ≥ 65 mph, 0 otherwise) | 0.41 | 0.49 | 0.43 | 0.50 | 0.56 | 0.50 |
| Lane1 (1 if number of lanes < 4, 0 otherwise) | 0.28 | 0.45 | 0.24 | 0.43 | 0.22 | 0.41 |
| Lane2 (1 if number of lanes ≥ 4, 0 otherwise) | 0.72 | 0.45 | 0.76 | 0.43 | 0.78 | 0.41 |
| AADT1 (1 if AADT ≤ 15,000, 0 otherwise) | 0.37 | 0.48 | 0.33 | 0.47 | 0.30 | 0.46 |
| AADT2 (1 if 15,000 < AADT ≤ 50,000, 0 otherwise) | 0.38 | 0.49 | 0.38 | 0.48 | 0.46 | 0.50 |
| AADT3 (1 if 50,000 < AADT ≤ 100,000, 0 otherwise) | 0.15 | 0.36 | 0.17 | 0.38 | 0.17 | 0.37 |
| AADT4 (1 if AADT > 100,000, 0 otherwise) | 0.10 | 0.29 | 0.12 | 0.33 | 0.07 | 0.25 |
| Asphalt (1 if asphaltic concrete surface, 0 otherwise) | 0.95 | 0.23 | 0.95 | 0.22 | 0.94 | 0.25 |
| Interstate (1 if interstate highway, 0 otherwise) | 0.50 | 0.49 | 0.54 | 0.50 | 0.62 | 0.49 |
| *Temporal characteristics* | | | | | | |
| Time1 (1 if time 7 AM–9:59 AM, 0 otherwise) | 0.17 | 0.38 | 0.17 | 0.37 | 0.18 | 0.38 |
| Time2 (1 if time 10 AM–3:59 PM, 0 otherwise) | 0.44 | 0.50 | 0.37 | 0.48 | 0.38 | 0.49 |
| Time3 (1 if time 4 PM–6:59 PM, 0 otherwise) | 0.16 | 0.37 | 0.16 | 0.37 | 0.12 | 0.32 |
| Time4 (1 if time 7 PM–6:59 AM, 0 otherwise) | 0.23 | 0.42 | 0.30 | 0.46 | 0.32 | 0.47 |
| Weekday (1 if weekday, 0 otherwise) | 0.89 | 0.32 | 0.87 | 0.33 | 0.78 | 0.42 |
| Weekend (1 if weekend, 0 otherwise) | 0.11 | 0.32 | 0.13 | 0.33 | 0.22 | 0.42 |

[†]SD = Standard Deviation

Variable descriptions and summary statistics by weather conditions are presented in Table 2. It should be noted that the HSIS database does not include all possible factors that contribute to injury severity of the truck drivers. Hence, the variables/factors considered in this study are limited to those available in the HSIS database.

## 4. Methodology

Mixed logit models are used to provide a better understanding of the interaction between crash factors found in the dataset and unobserved heterogeneity. Previous research has shown that models accounting for unobserved heterogeneity (i.e., mixed logit models) can be statistically superior. These models can account for observation-specific variations in the effects of explanatory variables. For that reason, mixed logit models are used more frequently in crash injury severity modeling (Anastasopoulos and Mannering, 2011; Anderson and Hernandez, 2017; Chen et al., 2019; Dong et al., 2018; Ma et al., 2015; Milton et al., 2008). The following subsections present the details of mixed logit modeling, estimation of marginal effects of the factors, and parameter transferability tests.

*4.1. Mixed logit model*

Following the methodology presented in previous research (i.e., Milton et al., 2008; Islam et al., 2014; Uddin and Huynh, 2017), the relationship between the injury severity variable and the explanatory variables is expressed as shown in Eq. (1).

$$Y_{in} = \beta_i X_{in} + \epsilon_{in} \tag{1}$$

where $Y_{in}$ is the variable representing injury severity level $i$ ($i \in I$ denotes injury severity levels, i.e., major injury, minor injury and no injury) of a truck driver $n$, $X_{in}$ is the injury severity explanatory variables/factors, $\beta_i$ is the parameter to be estimated for each injury severity level $i$, and $\epsilon_{in}$ is the error term to capture the effects of the unobserved characteristics. If the error term is independently and identically distributed with generalized extreme value distribution, then the resulting model is a multinomial logit model with the choice probability as shown in Eq. (2).

$$P_n(i) = \frac{\exp[\beta_i X_{in}]}{\sum_{i \in I} \exp[\beta_i X_{in}]} \tag{2}$$

where $P_n(i)$ is the probability of injury severity level $i$ for driver $n$.

To capture the effects of unobserved heterogeneity due to randomness associated with some of the factors necessary to understand injury sustained by the drivers, the above choice probability is extended to the mixed logit model formulation as shown in Eq. (3) (Train, 2009).

$$P_n(i|\phi) = \int \frac{\exp[\beta_i X_{in}]}{\sum_{i \in I} \exp[\beta_i X_{in}]} f(\beta_i|\phi) d\beta_i \tag{3}$$

where $P_n(i|\phi)$ is the probability of injury severity level $i$ conditional on $f(\beta_i|\phi)$, $f(\beta_i|\phi)$ is the density function of $\beta_i$ and $\phi$ is the parameter vector with known density function. Eq. (3) accounts for variations of the effects of the factors $X_{in}$, related to a specific injury severity level, in truck-involved crash probabilities for each weather condition model, where $\beta_i$ is determined using the density function $f(\beta_i|\phi)$. The mixed logit probabilities are calculated using weighted average for different values of $\beta_i$ across observations. Typically, some elements of $\beta_i$ are fixed and some are randomly distributed with specific statistical distribution. If the variance of $\phi$ is statistically significant, then the modeled injury severity levels vary with respect to $X$ across observations (Washington et al., 2011). In this study, maximum likelihood estimation is performed through a simulation-based approach to overcome the computation complexity of estimating the parameters $\beta_i$ of the mixed logit models. The simulation procedure requires Halton draws (Halton, 1960).

To test the overall model fit, the pseudo R-squared ($\rho^2$) value is used and is calculated using Eq. (4).

$$\rho^2 = 1 - LL(\beta)/LL(0) \qquad (4)$$

where $LL(0)$ is the log-likelihood at zero and $LL(\beta)$ is the log-likelihood at convergence.

*4.2. Marginal effects*

To determine the effect of a change in explanatory variable on the probability of injury severity level, marginal effects are calculated. The marginal effects for indicator variables are computed, as the difference in the estimated probabilities when the indicator variables change from 0 to 1, as shown in Eq. (5). Note that the marginal effects measure the discrete change (i.e., how predicted probabilities change as the explanatory variable changes from 0 to 1).

$$M_{X_{ink}}^{P_{in}} = P_{in}[\text{given } X_{ink} = 1] - P_{in}[\text{given } X_{ink} = 0] \qquad (5)$$

where $P_{in}$ is the probability of injury severity level $i$ for driver $n$ (i.e., Eq. (3)) and $X_{ink}$ is the $k$-th explanatory variable associated with injury severity level $i$ for driver $n$.

*4.3. Model separation*

Two different tests were conducted to validate that three separate weather condition models, one for each type of weather condition, is necessary. The first test is the log-likelihood ratio ($LR$) test between the full model and the weather condition models as shown in Eq. (6) (Washington et al., 2011).

$$LR_{full} = -2[LL(\beta^{full}) - LL(\beta^{normal}) - LL(\beta^{rain}) - LL(\beta^{snow})] \tag{6}$$

where $LL(\beta^{full})$ is the log-likelihood at convergence for the full model, $LL(\beta^{normal})$ is the log-likelihood at convergence for the normal condition model, $LL(\beta^{rain})$ is the log-likelihood at convergence for the rain model, and $LL(\beta^{snow})$ is the log-likelihood at convergence for the snow model. Note that log-likelihood values of the weather condition models have the same variables and specification as the full model. The $LR$ statistic is $\chi^2$ distributed, with degrees of freedom ($df$) equal to the summation of the number of estimated parameters in all three models minus the number of estimated parameters in the full model.

The second test is the parameter transferability test articulated in Washington et al. (2011). It is based on the $LR$ test and is used to determine if weather conditions are to be modeled separately. Its test statistic is computed using Eq. (7).

$$LR_{a_b} = -2[LL(\beta^{a_b}) - LL(\beta^a)] \tag{7}$$

where $LL(\beta^{a_b})$ is the log-likelihood at convergence of weather condition model $a$ using the data from model $b$ and $LL(\beta^a)$ is the log-likelihood at convergence of model $a$. The above test statistic has $df$ equals to the number of estimated parameters in $\beta^{a_b}$.

## 5. Results

The statistical software NLOGIT version 5 was used to perform the tests for model separation and to estimate the mixed-logit models (Econometric Software, Inc., 2019). The log-likelihood ratio test yielded a test statistic of 801.78 with 26 degrees of freedom ($p < 0.001$). These values suggest that weather condition models should be modeled separately with over 99% confidence. Subsequently, the parameter transferability test was conducted. Table 3 shows the results of this test. Each test statistic and its corresponding degrees of freedom suggest that weather condition models for truck-involved crashes should be modeled separately with well over 99% confidence.

**Table 3**
Test statistics, degrees of freedom and *p*-value of parameter transferability test.

| *a* | *b* | | |
| --- | --- | --- | --- |
| | Normal | Rain | Snow |
| Normal | 0 | 51.46, *df* = 15 ($p < 0.001$) | 38.09, *df* = 15 ($p < 0.001$) |
| Rain | 414.12, *df* = 10 ($p < 0.001$) | 0 | 32.92, *df* = 10 ($p < 0.001$) |
| Snow | 664.08, *df* = 13 ($p < 0.001$) | 37.30, *df* = 13 ($p < 0.001$) | 0 |

A separate model was estimated for each weather condition: normal, rain and snow. Each model predicts three levels of injury severity: major injury, minor injury and no injury. A simulation-based maximum likelihood method was utilized to estimate parameters $\beta_i$ for the mixed logit models. To estimate random parameters, the Normal, Lognormal, Triangular and Uniform distributions were considered. Only the Normal distribution was found to be statistically significant. This finding is consistent with previous studies where random parameters were considered (e.g., Milton et al., 2008; Morgan and Mannering, 2011; Behnood and Mannering, 2017a,b). Hence, the Normal distribution was used in the random parameters model. In addition, 500 Halton draws were utilized in the simulation procedure. During the model development process, variables were retained in the specification if they have *t*-statistics corresponding to the 90% confidence level or higher on a two-tailed *t*-test. The random parameters were retained if their standard deviations have *t*-statistics corresponding to the 90% confidence level or higher. Model estimation results are presented in Tables 4 through 6 along with marginal effects of all the variables included in the models. Note that only two constant terms can be used in the models since there are three injury severity levels. The estimation results yielded a 0 for one of the two constant terms used in the model specification. Other studies which performed similar analyses also reported having a 0 coefficient for one of the constant terms (e.g., Pahukula et al., 2015; Behnood and Mannering, 2015). For the above reason, there is only one constant term in the final estimated models under three weather conditions.

The $\rho^2$ values in Tables 4 through 6 indicate very good overall model fit with the values exceeding 0.60 in all three models. A total of 5 parameters were found to be statistically significant as random parameters among the three estimated mixed logit models. All of these random parameters were shown to be significantly different from zero with at least 90% confidence. These random variables account for unobserved heterogeneity.

Table 4 shows the model estimation results for crashes under normal condition. A positive coefficient value for an explanatory variable means it is positively associated with the injury severity level and increases the propensity of injury severity level with an increase in its magnitude. However, random variable results (mean and standard deviation) have a different interpretation. They indicate that one portion of the observations may have a higher probability of an injury severity level while the rest of the observations have a lower probability of that injury severity level, and vice-versa. For example, the parameter *weekend* (specific to minor injury) was found to be random and had a mean of −1.91 and standard deviation of 2.54. With these values, the Normal distribution curve indicates that 77.4% of the truck-involved crashes that occurred during the daylight under normal condition had a higher probability of drivers sustaining a minor injury, while the rest (100 − 77.4 = 22.6%) of the crashes had a lower probability of drivers sustaining a minor injury. In the following, results of the random parameters are reported without the mean and standard deviation. Also, statements regarding the "rest of the crashes" are omitted since they can be deduced from the reported findings.

**Table 4**
Parameter estimates and marginal effects for truck-involved crashes under normal condition.

| Variable | Coefficient | $t$-statistic | $p$-value | Marginal effects | | |
|---|---|---|---|---|---|---|
| | | | | Major injury | Minor injury | No injury |
| *Defined for major injury* | | | | | | |
| Male | −4.49 | −44.53 | 0.000 | −0.048 | 0.004 | 0.044 |
| Rear-end | −0.58 | −5.43 | 0.000 | −0.003 | 0.000 | 0.003 |
| Dark-lighted | 0.42 | 3.51 | 0.001 | 0.001 | −0.000 | −0.001 |
| Time1 | 0.71 | 7.21 | 0.000 | 0.004 | −0.000 | −0.004 |
| | | | | | | |
| *Defined for minor injury* | | | | | | |
| Constant | −1.30 | −10.06 | 0.000 | | | |
| Rural | 1.24 | 15.32 | 0.000 | −0.000 | 0.011 | −0.011 |
| Single-unit truck | −0.51 | −4.88 | 0.000 | 0.000 | −0.026 | 0.026 |
| Lane2 (standard deviation of parameter distribution) | −1.90 (3.67) | −5.09 (9.00) | 0.000 (0.000) | −0.000 | 0.020 | −0.020 |
| Asphalt | −0.74 | −9.30 | 0.000 | 0.001 | −0.035 | 0.034 |
| Weekend (standard deviation of parameter distribution) | −1.91 (2.54) | −7.42 (10.52) | 0.000 (0.000) | −0.000 | 0.021 | −0.021 |
| *Defined for no injury* | | | | | | |
| Sideswipe | −1.07 | −15.53 | 0.000 | 0.003 | 0.015 | −0.018 |
| Object | 0.31 | 4.80 | 0.000 | −0.001 | −0.004 | 0.005 |
| Speed2 | 1.52 | 13.95 | 0.000 | −0.001 | −0.008 | 0.009 |
| *Model statistics* | | | | | | |
| Number of observations | 40,459 | | | | | |
| Log-likelihood at zero, $LL(0)$ | −44,448.77 | | | | | |
| Log-likelihood at convergence, $LL(\beta)$ | −14,687.87 | | | | | |
| $\rho^2 = 1 - LL(\beta)/LL(0)$ | 0.67 | | | | | |

The other significant random parameter for the normal condition model is *lane2*. Specific to minor injury, about 69.8% of the crashes occurring on 4 or more lanes (both directions) highway under normal condition had a higher probability of drivers sustaining a minor injury.

**Table 5**
Parameter estimates and marginal effects for truck-involved crashes under rainy condition.

| Variable | Coefficient | *t*-statistic | *p*-value | Marginal effects | | |
|---|---|---|---|---|---|---|
| | | | | Major injury | Minor injury | No injury |
| *Defined for major injury* | | | | | | |
| Male | −1.70 | −10.96 | 0.000 | −0.032 | 0.002 | 0.030 |
| Speed3 | 0.70 | 2.38 | 0.018 | 0.004 | −0.001 | −0.003 |
| *Defined for minor injury* | | | | | | |
| Sideswipe | 0.92 | 3.72 | 0.000 | −0.000 | 0.010 | −0.010 |
| Single-unit truck (standard deviation of parameter distribution) | −2.94 (3.35) | −2.83 (3.60) | 0.005 (0.000) | −0.000 | 0.033 | −0.033 |
| Interstate | −0.50 | −1.75 | 0.081 | 0.000 | −0.003 | 0.003 |
| Weekend | −0.69 | −3.20 | 0.001 | 0.000 | −0.020 | 0.020 |
| *Defined for no injury* | | | | | | |
| Constant | 1.90 | 8.12 | 0.000 | | | |
| Rural | −1.26 | −5.54 | 0.000 | 0.004 | 0.014 | −0.018 |
| Daylight | −0.36 | −2.18 | 0.029 | 0.001 | 0.006 | −0.007 |
| *Model statistics* | | | | | | |
| Number of observations | 4,866 | | | | | |
| Log-likelihood at zero, $LL(0)$ | −5,345.85 | | | | | |
| Log-likelihood at convergence, $LL(\beta)$ | −1,844.82 | | | | | |
| $\rho^2 = 1 - LL(\beta)/LL(0)$ | 0.65 | | | | | |

Table 5 shows the model estimation results for crashes under rainy condition. The significant random parameter for the rainy condition model is *single-unit truck*. Specific to minor injury, about 81.0% of the crashes involving single-unit trucks under rainy condition had a higher probability of drivers sustaining a minor injury.

**Table 6**
Parameter estimates and marginal effects for truck-involved crashes under snowy condition.

| Variable | Coefficient | t-statistic | p-value | Marginal effects | | |
| --- | --- | --- | --- | --- | --- | --- |
| | | | | Major injury | Minor injury | No injury |
| *Defined for major injury* | | | | | | |
| Curve (standard deviation of parameter distribution) | 0.78 (1.53) | 2.42 (1.75) | 0.016 (0.080) | 0.003 | −0.000 | −0.003 |
| Single-unit truck | −1.66 | −3.77 | 0.000 | −0.013 | 0.001 | 0.012 |
| Time3 | 0.59 | 1.79 | 0.073 | 0.003 | −0.000 | −0.003 |
| | | | | | | |
| *Defined for minor injury* | | | | | | |
| Male (standard deviation of parameter distribution) | −1.32 (2.00) | −1.42 (2.47) | 0.155 (0.014) | −0.001 | 0.087 | −0.086 |
| Truck trailer | −1.05 | −1.93 | 0.054 | 0.000 | −0.007 | 0.007 |
| Interstate | 1.08 | 1.92 | 0.055 | −0.001 | 0.009 | −0.008 |
| | | | | | | |
| *Defined for no injury* | | | | | | |
| Constant | 4.53 | 8.15 | 0.000 | | | |
| Urban | −0.99 | −3.19 | 0.001 | 0.006 | 0.016 | −0.022 |
| Rear-end | −1.16 | −3.57 | 0.000 | 0.008 | 0.024 | −0.032 |
| Object | 1.11 | 2.70 | 0.007 | −0.002 | −0.011 | 0.013 |
| | | | | | | |
| *Model statistics* | | | | | | |
| Number of observations | 3,923 | | | | | |
| Log-likelihood at zero, $LL(0)$ | −4,309.86 | | | | | |
| Log-likelihood at convergence, $LL(\beta)$ | −1,594.62 | | | | | |
| $\rho^2 = 1 - LL(\beta)/LL(0)$ | 0.63 | | | | | |

Table 6 shows the model estimation results for crashes under snowy condition. The significant random parameters for the snowy condition model are *curve* and *male*. Specific to major injury, about 30.5% of the crashes occurred on curved highway segment under snowy condition had a higher probability of drivers sustaining major injury. Specific to minor injury, about 74.5% of the crashes where drivers were male under snowy condition had a higher probability of sustaining minor injury.

## 6. Discussion

Separate models of injury severity levels by weather conditions provide valuable insights about contributing factors affecting the injury severity of truck-involved crashes. The model results suggest major differences in both the combination and magnitude of impact of variables. For example, single-unit truck drivers were found to be associated with decreased probability of minor injury under normal condition, increased probability of minor injury under rainy condition and decreased probability of major injury under snowy condition. Some variables were found to be significant in one weather condition but not in others. For example, the *curve* variable is only significant in contributing to major injury under snowy condition. Table 7 compares the effects of the statistically significant factors on injury severity by weather conditions.

*6.1. Driver characteristics*

Male drivers were found to have lower probability of major injuries under normal and rainy conditions; however, they were found to have higher probability of minor injury under snowy condition. Specifically, compared to female drivers, the probability of sustaining a major injury by male drivers was lower by 0.048 under normal condition and 0.032 under rainy condition. Under snowy condition, compared to female drivers, the probability of sustaining a minor injury by male drivers was higher by 0.087. This indicates that male drivers were less likely to sustain a severe injury compared to female drivers. This finding is consistent with those reported in O'Donnell and Connor (1996).

**Table 7**
Model comparisons.

| Variable | Normal | | | Rain | | | Snow | | |
|---|---|---|---|---|---|---|---|---|---|
| | Major | Minor | No | Major | Minor | No | Major | Minor | No |
| Male | ⇩ | | | ⇩ | | | | ⇧ | |
| Rural | | ⇧ | | | | ⇩ | | | |
| Urban | | | | | | | | | ⇩ |
| Curve | | | | | | | ⇧ | | |
| Rear-end | ⇩ | | | | | | | | ⇩ |
| Sideswipe | | ⇩ | | | ⇧ | | | | |
| Object | | | ⇧ | | | | | | ⇧ |
| Daylight | | | | | | ⇩ | | | |
| Dark-lighted | ⇧ | | | | | | | | |
| Single-unit truck | | ⇩ | | | ⇧ | | ⇩ | | |
| Truck trailer | | | | | | | | ⇩ | |
| Speed2 | | | ⇧ | | | | | | |
| Speed3 | | | | | ⇧ | | | | |
| Lane2 | | ⇧ | | | | | | | |
| Asphalt | | ⇩ | | | | | | | |
| Interstate | | | | | ⇩ | | | ⇧ | |
| Time1 | ⇧ | | | | | | | | |
| Time3 | | | | | | | ⇧ | | |
| Weekend | | ⇧ | | | ⇩ | | | | |

⇧ indicates increase and ⇩ indicates decrease in the probability of an injury severity level.

## 6.2. Crash characteristics

Crashes occurring in rural areas were found to increase the probability of minor injury by 0.011 under normal condition and decrease the probability of no injury by 0.018 under rainy condition. On the other hand, crashes occurring in urban areas were found to decrease the probability of no injury by 0.022 under snowy condition. A possible reason for this finding is that crashes occurring in rural areas increase the chance of minor injury, but since drivers are more cautious during rainy and snowy conditions their chance of sustaining an injury is low. Crashes occurring on horizontal curves were found to increase the probability of major injury by 0.003 under snowy condition. Similar results have been reported by Anderson and Hernandez (2017), Islam et al. (2014), Naik et al. (2016) and Osman et al. (2016). For example, Anderson and Hernandez (2017) found that horizontal curves increase the probability of injury on U.S. and state highways.

Rear-end crashes were found to decrease the probability of major injury by 0.003 under normal condition and decrease the probability of no injury by 0.032 under snowy condition. A possible explanation is that under normal condition when a truck is struck from behind by another vehicle, it is less likely to cause a major injury for the driver. Sideswipe collisions were found to decrease the probability of no injury by 0.018 under normal condition and increase the probability of minor injury by 0.010 under rainy condition. One possible reason could be when a truck is involved in sideswipe collision with another vehicle under normal condition it is less likely to cause major or minor injury for the driver. However, under rainy condition, the sideswipe collision may cause the vehicle to stray from its lane or road, and thus, resulting in a higher probability for minor injury. Hitting an object was found to increase the probability of no injury by 0.005 under normal condition and increase the probability of no injury by 0.013 under snowy condition. This result is consistent with the finding of Naik et al. (2016), where it is reported that hitting fixed objects are associated with less severe injuries.

Two of the lighting condition variables were found to be significant: daylight and dark-lighted. Crashes under daylight were found to decrease the probability of no injury by 0.007 under rainy condition. Furthermore, crashes under dark with streetlights were found to increase the probability of major injury by 0.001 under normal condition. One possible explanation for truck drivers experiencing higher probability of major injury could be the poor visibility under rainy condition during nighttime. This finding suggests that roadway visibility has significant impact on the driver injury severity. Previous truck-involved crash studies reported the similar findings as well (e.g., Pahukula et al., 2015; Uddin and Huynh, 2017).

*6.3. Vehicle characteristics*

Single-unit trucks were found to decrease the probability of minor injury by 0.026 under normal condition, increase the probability of minor injury by 0.033 under rainy condition and decrease the probability of major injury by 0.013 under snowy condition. This finding suggests that single-unit truck drivers are less likely to experience severe injuries from crashes under normal and rainy condition; however, they are more likely to experience severe injuries from crashes under snowy condition. This may be due to the combined effects of trucks being heavy and slippery road conditions due to snow, which makes harder to stop and easier to lose control. Truck trailers were found to decrease the probability of minor injury by 0.007 under snowy condition. A possible explanation could be the drivers being more cautious under snowy condition.

*6.4. Roadway characteristics*

Speed limit being 45 to 60 mph was found to increase the probability of no injury by 0.009 under normal condition. Speed limit being 65 mph or higher was found to increase the probability of major injury by 0.004 under rainy condition. This finding suggests that higher speed limits have a potential adverse effect on truck safety. The finding is consistent with those reported in previous studies (e.g., Cerwick et al., 2014; Chang and Mannering, 1999; Chen et al., 2018; Uddin and Huynh, 2017). Number of lanes being 4 or more was found to increase the probability of minor injury by 0.020 under normal condition. Asphaltic concrete surface was found to decrease the probability of minor injury by 0.035 under normal condition. Interstate highway was found to decrease the probability of minor injury by 0.003 under rainy condition and increase the probability of minor injury by 0.009 under snowy condition. A possible explanation is that during inclement weather condition, truck drivers are more cautious. The combination of the drivers being more cautious and slower vehicle speed reduces the risk of severe injury.

*6.5. Temporal characteristics*

Crashes occurring during the morning peak hours (7 to 9:59 AM) were found to increase the probability of major injury by 0.004 under normal condition. In addition, crashes occurring during the evening peak hours (4 PM to 6:59 PM) were found to increase the probability of major injury by 0.003 under snowy condition. This is perhaps because of the combined effects of severe collisions due to high traffic volume and dark lighting condition in the fall and winter. Weekend crashes were found to increase the probability of minor injury by 0.021 under normal condition and decrease the probability of minor injury by 0.020 under rainy condition. A possible explanation is that traffic volume tends to be lower on commuter routes on the weekends, and thus, crashes resulting in less severe injury.

## 7. Conclusion

This study investigated truck driver injury severity under different weather conditions using crash data from the state of Ohio from 2011 to 2015. Two likelihood ratio tests were conducted to test the hypothesis that separate models are warranted for different weather conditions. The results of these tests suggested that separate weather condition models are needed, particularly those in the HSIS database. Subsequently, three weather condition models were estimated: normal, rain and snow. A good number of the statistically significant variables were found to be exclusive to each weather condition model, which further underscores the need to examine driver injury severity in truck-involved crashes for different weather conditions. Specifically, it was found that 5 significant variables were exclusive to crashes during normal condition (dark-lighted, speed limit (45 to 60 mph), 4 or more lanes, asphalt, and time (7 AM to 9:59 AM)), 3 significant variables were exclusive to crashes during rainy condition (daylight, speed limit (≥ 65 mph), and interstate), and 4 significant variables were exclusive to crashes in snowy condition (urban, curve, truck trailer, and time (4 PM to 6:59 PM)). The parameters *male* and *single-unit truck* were found to have an impact on driver injury severity across all weather conditions. Rural, rear-end, and sideswipe crash parameters were found to have significantly different levels of impact on injury severity in truck-involved crashes.

The results obtained from this study's developed models have a number of implications. First, male drivers were found to sustain less severe injuries compared to female drivers. This finding suggests that safety and enforcement programs should focus on female truck drivers; perhaps, providing additional driving training and/or traffic safety course. They should be taught to obey the traffic rules and regulations strictly to improve their safety while driving. Second, higher speed limit was found to be positively associated with major injuries under rainy condition. This finding suggests that the use of a variable speed limit sign to lower speeds during rainy condition may reduce injury severity in truck-involved crashes. Third, it was found that truck drivers were less likely to sustain severe injuries on interstates under both rainy and snowy conditions. This finding suggests that trucks should be restricted or prohibited on certain non-interstate routes under rainy and snowy conditions. Lastly, during afternoon peak (4 PM to 6:59 PM) under snowy condition, it was found that truck drivers were more likely to be involved in major injuries. A possible explanation is that the evening rush hour could lead to aggressive driving. This finding suggests that supply chains and logistics policies should be put in place to allow trucks to make deliveries during off peak hours.

Similar to most past studies, this study has the limitation of using crash data from a single state. This fact should be taken into account when interpreting and applying the findings. In future research, it would be more insightful if researchers were to combine crash data from multiple states and different databases. In a few recent studies, it has been reported that crash data may have temporal instability due to a number of fundamental behavioral reasons (Behnood and Mannering, 2015, 2019; Mannering, 2018). It is technically challenging to explicitly account for temporal elements based on the current modeling approaches according to Mannering (2018). This is a new area of research that could potentially lead to a new paradigm for modeling crashes.

**Acknowledgment**

The authors would like to thank Ms. Anusha Patel Nujjetty from the Highway Safety Information System laboratory for providing the data used in this study.